\documentclass[useAMS,usenatbib,onecolumn,referee]{mn2e}
\usepackage{graphicx}

\title[Composition of the Grain Mantle]
{Effects of Initial Condition and Cloud Density on the Composition of the Grain Mantle}
\author[Ankan Das, Kinsuk Acharyya, Sandip K. Chakrabarti]
{Ankan Das$^{1,2}$, Kinsuk Acharyya$^{2}$, Sandip K. Chakrabarti$^{2}$\\
$^{1}$Indian Centre for Space Physics, Chalantika 43, Garia Station Rd.,
             Kolkata, 700084, India\\
$^{2}$S. N. Bose National Centre for Basic Sciences, Salt Lake,
              Kolkata 700098, India}
\begin{document}

\date{}


\maketitle


\begin{abstract}
Evolution of grain mantles in various interstellar environment is studied. We 
concentrate mainly on water, methanol, carbon di-oxide,
which constitute nearly $90 \%$ of the grain mantle. We investigate how the production 
rates of these molecules depend on the relative gas phase abundances 
of oxygen and carbon monoxide and constrain the relevant parameter space 
which reproduces these molecules closed to the observed abundances.
Allowing to accrete only H, O and CO on the grains and using the Monte-Carlo method 
we follow the chemical processes for a few million years. We allow formation of multi-layers
on the grains and incorporate the freeze-out effects of accreting O and CO. 
We find that the formation of these molecules depends
on the initial conditions as well as the average cloud density.
Specifically, when the number density of accreting O is less than $3$ times more than that 
of CO, methanol is always over-produced. Using available reaction pathways 
it appears to be difficult to match the exact observed abundances of all the three molecules 
simultaneously. Only in a narrow region of parameter space all these three
molecules are produced within the observed limit. In addition to this, we found that the
incorporation of the freeze-outs of O and CO leads to almost steady 
state on the grain surface. The mantle thickness grows anywhere between $60$ to $500$ layers 
in a period of two million years. In addition, we consider a case where the gas number density changes 
with time due to gradual collapse of the molecular cloud and present the evolution of composition
 of different species as a function of radius of the collapsing cloud.
\end{abstract}


\section{Introduction}
\noindent The study of chemical evolution of interstellar medium is well recognized to 
be a challenging task. Interstellar medium (ISM) is a rich reservoir of complex molecules. 
So far, around $150$ gas phase molecules and 
around $20$ molecular species on the grain surface have been detected in the various 
regions of ISM, especially in the regions of star formation. In the last decade, it was 
well established that the gas phase reactions alone cannot explain the 
molecular abundances in the ISM. The chemical reactions which occur on the interstellar 
dust grains are essential to explain the formation of several molecules especially 
hydrogenated species including the simplest and the most abundant 
molecule H$_2$. Interstellar grains provide the surface for the accreted species to 
meet and react. Therefore, an understanding of the formation of molecules on the grain 
surfaces is of prime importance. In this paper, we mainly 
follow the evolution of the grain mantle as a function of the initial condition. We 
concentrate only on water, methanol and carbon di-oxide, since they constitute nearly 
$90 \%$ of the grain material in dense regions of ISM.  These molecules are detected 
on the grain surface due to their strong absorption bands arising out of multiple vibrational modes.
Water is the most abundant species on a grain in the dense interstellar medium. 
It has an abundance of $10^{-4}$ with respect to the total hydrogen column density 
\citep{Tiel91}. The ice band, at $3.07~\mu$m ($3280$ cm$^{-1}$) is due to the O-H stretch 
mode of H$_2$O ice and it was first discovered by \citet{Gill} in BN objects. CO$_2$ is 
the second most abundant molecule in the interstellar medium with an abundance of around 
$20 \%$ with respect to H$_2$O. However, this abundance can vary 
from cloud to cloud and in clouds like W 33A it could be even less than $5 \%$ of 
water abundance \citep{Kean}. Several strong CO$_2$ features are observed \citep{Degr,Kean}. 
Generally, it is found that the CO$_2$ correlates best with H$_2$O ice, 
revealing the fact that these molecules may have similar chemical history. 
The next most abundant molecule is CO, which is the well studied ice with an abundance 
varying between $2\%$ to $15\%$ of water and with a characteristic absorption feature 
near $2140$ cm$^{-1}$($4.67$ $\mu$m). The next strong feature is due to methanol 
(CH$_3$OH), its abundance could vary between $2 \%$ to $30 \%$ of water. However, in 
most of the cloud it is less than $10 \%$. Observed abundances around a few clouds
are used from \cite{Kean} and references therein.

\citet{Holl}, first introduced the grain surface chemistry to explain the formation of 
molecular hydrogen. Since then it has 
been used very widely by several authors \citep{Wata,Watb,Al75,Al76,Al77,Tiel82,Hase,
Charn,Stan,Acha05,Biha,Chaka,Green,Dasa,Dasb}. These studies mainly belong to two different 
categories, namely, the deterministic approach and the stochastic approach. In the 
deterministic approach, one can completely determine the time evolution of the system 
once the initial conditions are known. Rate equation method belongs to this category. 
This method is very extensively used by several authors to study the grain surface 
chemistry \citep{Herb73,Pras80,Hase,Robe}. However, this method is only applicable 
when there are large number of reactants on the grain surface. Given the fact that 
the interstellar medium is very dilute, very often this criteria is not fulfilled and this 
method cannot be applicable. But this method is computationally faster and can very 
easily be coupled with the gas phase reactions. In the stochastic approach, fluctuations 
in the surface abundance due to the statistical nature of the grain is preserved. The Monte 
Carlo method and the Master equation methods belong to this category. Both these 
methods are used by several authors \citep{Charn,Green,Biha,Stan}. 
Its major disadvantage is that it is computationally intensive.
Coupling of Monte-Carlo method to study grain surface 
reactions and rate equation method for gas phase reactions is extremely difficult. 
The Master equation method can be coupled with the rate 
equations, however, it is disadvantageous because for a large network one had to 
solve a large number of reactions 

Recently, \citet{Chang} argued that the stochastic methods used so far can also 
lead to error because the rate of reaction is determined by the rate of hopping 
(or tunneling) of a hydrogen atom from one site to the nearest neighboring site 
multiplied by the probability of finding a reactant partner in this site. This is 
also an average treatment since on any given grain the reactant partner is unlikely to 
lie in the nearest-neighboring site. They used a continuous random work technique to 
study the formation of molecular hydrogen. \citet{Chaka,Chakb} used a similar method 
which keeps track of each individual reactant and their movements and calculated effective grain 
surface area involved in the formation of molecular hydrogen in the interstellar clouds. 
Furthermore, they defined an important parameter called `catalytic capacity' which measures 
the efficiency of the formation of H$_2$ on a grain surface for a given pair of H residing on 
it. They studied the formation of water and methanol up to a mono-layer and estimated 
the abundances of these two molecules. They found that the formation rate of various 
molecules is strongly dependent on the binding energies, number density in the gas 
phase, effective grain surface area and on the formation mechanism \citet{Dasb}.

In this we paper, we first vary the elemental abundances of O and CO and found regions in which 
these molecules are produced efficiently. We then changed the cloud density and found how the
abundance is affected. In Section 2, we discuss
various physical processes that are involved during a gas-grain interaction. 
In Section 3, we describe the methodology of our calculations. In Section 4, we 
describe our model and the nature of the initial conditions.
In Section 5, the results are presented. Finally, in Section 6, we draw our conclusions.

\section{Mechanisms of reactions on grain surfaces}

There are four physical processes that are involved while gas phase atoms/molecules 
interact with the grains. The first step is `accretion', i.e., landing of various species 
onto a grain. In our case only H, O and CO are considered as
accreting species onto the grain surface. The second step is `hopping' through which the accreted species 
move around the grain. In the third step, the accreted species will react to form various new species
either through the Langmuir-Hinselwood (LH) mechanism or by the Eley-Rideal (ER) mechanism. 
In the LH scheme, the gas phase species accreted onto a grain becomes equilibrated 
with the surface before it reacts with another atom/molecule, and in the ER reaction scheme, 
incident gas phase species collides directly with an adsorbed species on the surface
and reacts with that species. In such a mechanism, generally the reactant does not become 
trapped at the surface and it is unlikely to be sensitive to the surface temperature 
(Farebrother et al. 2000). In our study, we take it that the reactions on a surface 
can occur through both LH and ER mechanisms. However, we assume that the molecules remain
trapped after reactions due to their high binding energy. Typical temperature of dense 
cloud is around 10 K, and in this temperature only hydrogen 
can desorb at a meaningful rate. All the other molecules stay on the grain until 
these grains are heated up.
When the temperature of the grain is increased, these molecules desorb back into the gas 
phase according to their binding energy of desorption.

We define the accretion rate [$r_{acc}(i)$] of a given neutral species $i$, in the 
units of s$^{-1}$ , as,
\begin{equation}
r_{acc} (i) = S_i \sigma v_i n_i,  
\end{equation}
where, $S_i$ is the sticking coefficient (taken as $1$ for all the three species), $v_i$ 
is the velocity (cm s$^{-1}$ ), $n_i $ is the number density of the $i$-th species in the 
gas phase and $\sigma$ is the grain cross-section (cm$^2$). Since 
we have carried out our calculations for three different cloud densities and different 
initial abundances we have used many sets of accretion rates. Number densities for 
different clouds are taken from \citet{Stan} and are listed in Table 1. 
Number density of hydrogen in gas phase is denoted by $n_h$ and the number density 
of O and CO with respect to hydrogen number density is expressed as $n_O$ and $n_{CO}$
respectively.

The binding energy of the incoming species greatly depends on the species itself 
and on the way the interactions proceeded. The incoming species might get trapped inside 
a shallow potential well at a physisorbed site. The interaction is mainly due to mutually 
induced dipole moments. A strong covalent bond may also be formed through chemisorption. Recent 
studies have found the evidences of both physisorption and chemisorption processes \citet{Caz02}. 
However, for the chemisorption, high kinetic energy is involved and hence this type of interaction 
is not relevant except in a very special astrophysical conditions. Therefore, we  
considered only weakly bound species, i.e., physisorbed atoms and molecules. 
The typical energy for physisorption is around $0.1$ eV or $ \sim \ 1000$ K. Let us denote 
the binding energy for physical adsorption by $E_D$ and, $E_b$, the potential energy 
barrier which must be tackled in order that the species diffuses from 
one site to other. We need many interaction energies to describe the system completely, 
not just among the silicate surfaces and adsorbed species but also among the adsorbed 
species themselves. Table 2, contains all the binding energies that are used 
in our calculation. However, binding energies are not known with certainty. We took 
the binding energies of H and H$_2$ on the silicate surface from \citet{Katz99}, the
binding energies of other species with bare silicate grain from \citet{Al77,Tiel87,Hase}, 
and the binding energies among H, O, OH and H$_2$O from \citet{Cup07}. For the rest,
we kept the energies same as those on the silicate surface. These are taken from \cite{Hase93}.

For a chemical reaction to occur, an accreted species has to scan the surface in search 
of a reaction partner. There are two physical processes which can provide the mobility 
for the accreted gas phase species, namely, the thermal hopping and tunneling. 
As mentioned earlier, \citet{Holl} first introduced the grain surface chemistry to explain 
the high abundance of molecular hydrogen. They assumed that within the grains, hydrogen atoms 
move from site to site by quantum mechanical tunneling process. But 
from the recent experimental results of molecular hydrogen formation it was found that the 
mobility of hydrogen on grains is primarily due to thermal hopping \citep{Pir97a,Pir97b,Pir99}. 

In the paper, we have considered the thermal desorption only. However, other than the hydrogen 
and hydrogen molecule no species desorbs at $10$ K. 
Other molecules can come out of the grain surface only when the grains are heated up.
Many different types of desorption mechanisms are present in the literature. In the dense cloud,
the most important is impulsive heating of the grains (\citet{Hase93}) by the cosmic rays. 
However, this mechanism is effective for small molecules. Following \citet{Hase93},
and assuming the binding energies of the species with water, we obtain the time scales for CO, O$_2$, 
H$_2$O, CH$_3$OH and CO$_2$ are $10^{14}$ sec, $10^{14}$ sec, $3 \times 10^{41}$ 
sec, $2 \times 10^{19}$ sec and $10^{22}$ sec respectively. Thus the cosmic ray induced desorption 
mechanism is completely in-effective for H$_2$O, CH$_3$OH and CO$_2$ and weakly effective 
for CO and O$_2$ at much later time. 
In addition to this, 
\citet{Shen04} argued that \citet{Hase93} over-estimated the rate as 
they assumed that the sublimation of the volatile species such as CO occurs near 70K,
but it is not possible for the Cosmic ray particle to heat a bigger 
grain(due to dust accretion the size of the grain in fact increases) upto such
high temperature. Finally \citet{Shen04} concluded that
the CR induced UV field is about 10 times more efficient in depositing energy
in the ice than the direct CR energy deposition. So if we sacle our calculated time scale
of desorption accordingly, we can conclude that it is again in-effective for H$_2$O($\sim 3 \times 10^{40}$sec),
CH$_3$OH($\sim 2 \times 10^{18}$sec) and CO$_2$($10^{21}$sec). For CO and O$_2$ it is $\sim 10^{13}$sec $\sim 10^6$year, 
which is our simulation time scale. So we can conclude that inclusion of these
desorption mechanism will not significantly affect our results.
\citet{Rob07}, suggested 
a new type of desorption mechanism in which they used exothermicity 
of surface addition reaction to desorb product molecules from the surface. 
They used this mechanism parametrically for gas-grain interaction 
code and found that the maximum desorption due to this mechanism is 10 $\%$. It is 
important only at later times (after 3-4 Million years). Therefore, we have 
not included this type of desorption in our computation. 

Along with the physical processes one also needs to know what are the reaction pathways which 
can lead to the formation of these molecules. All the reaction pathways that are considered in our 
study are listed in Table 3. We assume that water is formed through H $+$ O $\rightarrow$ OH, 
H $+$ OH $\rightarrow$ H$_2$O (Eq. 2 and 3) \citep{Hase,Stan}. Recently, \citet{Ioppo}, 
suggested formation of water through H $+$ O$_2$ channel in which O$_2$ is converted to H$_2$O 
via H$_2$O$_2$. However, considering that water is the most abundant species on a
grain surface and there is very little observed O$_2$ and H$_2$O$_2$, we have not considered this 
path way. Methanol formation through hydrogenation is studied by three groups\citep{Hiroka,Watan,Fuchs}.
\citet{Hiroka} observed only formaldehyde formation, whereas \citet{Watan}
also found efficient methanol production. These two groups presented these apparently contradicting 
results in a series of papers and this discrepancy was mainly attributed to be a result of 
different experimental conditions, especially on H-atom flux. \citet{Fuchs}, 
found that the formation mechanism of formaldehyde and methanol does not fundamentally 
change with varying flux and concluded that the surface hydrogenation
of CO can safely be used to explain the majority of the formed methanol in the interstellar 
medium. We have also considered methanol formation through successive hydrogenation of 
CO through Eq. 4, 5, 6 and 7. Among these, Eqs. 4 and 6 have an activation barrier of $2000$ K. 
O$_2$ is formed through Eq. 9. Finally, CO$_2$ is formed through Eq. 9 and 10 \citet{Stan}.

\section{Procedure}
We have used continuous-time random walk Monte Carlo method to study the evolution of grain 
mantle. For the sake of simplicity, we take all the grains 
to be of square shaped and assume that each site has four nearest neighbors, as in an fcc[100] 
plane. In order to mimic the spherical grain structure, we assume periodic boundary 
condition i.e., a species which leaves the simulation grid from one boundary enters back from 
the opposite boundary. In the earlier Section, we discussed all the steps which occur on a grain 
surface and we implemented these steps in following way: In the first step, we drop atoms/molecules 
after every $1/r_{acc}$ seconds. Since we are considering three different species we have three
different accretion rates for each species depending on the number densities. Typically, 
for a cloud of intermediate number density and a classical grain it is $1.2 \times 10^5$ seconds
(order of a day), $7.5 \times 10^6$ seconds ( $\sim$ couple of months) and $8.75 \times 10^5$ 
seconds ($\sim$  weeks) for H, O and CO respectively at the beginning. 
The location of the accreting $i$-th species is dictated by a pair of random numbers 
($R_x, R_y$) obtained by a random number generator. This pair would place the incoming species 
at (j, k)$^{th}$ site of the grain, where j and k are the nearest integers obtained using Int 
function: $j=int(n R_x +0.5)$ and $k=int(n R_y+0.5)$. Here, $n = \sqrt s$, where, $s$ is the 
number of sites on the grain. We have considered $50 \times 50$ grain. It was found 
by \citet{Chang} that when the size of the simulated surface is at least $50 \times 50$ 
there is no size effect. We then scaled the result for classical grains ($\sim 10^6$ sites). 
While considering the grain mantle formation in multilayer 
regime, two possibilities may occur depending on whether the site is already occupied or not. 
If the site is not occupied, then the accreted species can sit on the surface of the 
bare silicate grain at the randomly generated (j, k)$^{th}$ location. But if the site 
is occupied then either a reaction between the incoming and the stationary species will 
occur, provided the reaction is permitted (ER mechanism) and will form a new molecule 
or it can diffuse through the grain according to the binding energy with species. In the 
second step, all the accreted species are allowed to hop. We choose
$1/a_h$, the hopping time of hydrogen as the minimum time step and advance the global 
time by this time step. Hopping time of H, O and CO with the silicate surface at $10$ K is 
$7.5 \times 10^{-7}$ second, $0.02$ seconds and $6000$ seconds respectively. Since, 
CO hopping time is much higher with respect to O and H, we have not 
considered the hopping of CO. 

It is generally observed that the hopping time scale is many order of magnitude smaller than the accretion 
time scale. Therefore, once one H or O lands on a grain surface, 
it can scan the grain very efficiently and react almost instantaneously due to the thermal hopping.
Tunneling also provides a mobility to hydrogen atoms but it is much faster than hopping. 
Therefore, inclusion of tunneling means that the scanning can proceed even faster. But as the accretion 
time scale is not altered, we will not have any major differences in our result. 
Already in \citet{Dasb}, we did not find any major difference between the results obtained with the
hydrogen tunneling and the hydrogen hopping. \citep{Pir97a,Pir97b,Pir99} did not find the experimental 
evidence of tunneling either. However, many models, e.g., \citet{Caz08}, still use tunneling for the mobility of hydrogen atoms.  

In an fcc[100] plane there are four directions to hop. We generate random numbers for the four nearest neighbor of the diffusing 
species to decide the direction (n, l) of it. Here we are considering all the situations by the Markov chain Monte Carlo method which is a 
random walk process and the walker will often move back and cover the place already covered. So by considering this we automatically include the 
back-diffusion probability mentioned by \citet{Chang}. Once again, two possibilities may occur: If the site is occupied then either it will react 
when the reaction is permitted or it will wait until the next hopping time. If the neighboring site is vacant and it does not have any species 
just beneath that site then it can roll down on that (n,l)$^{th}$ grid until it touch some species or the bare grain surface. If the reaction 
between the species and the diffusing species is permitted, a new species is formed and if not, 
then it can sit on the top of that. 
From Table 2, it is to be noted that some species are very weakly bound with the other surface species such as, O with 
O$_2$. Since, the hopping probability is higher than that of the evaporation, it is most likely that the species will drop down from the top of the weak 
binding species rather than evaporating from there. Finally, the species are desorbed back into the gas phase. This is also done by generating random 
numbers. Actually, after each hopping we generated a random number to decide if the species will hop or desorb. We have evolved our system 
up to a few million years, which is typical life time of a molecular cloud. 

In \citet{Dasb}, the simulations were restricted in the mono-layer regime. Therefore, it was not possible to study
the evolution and composition of the grain mantle with time. In this paper, 
we used binding energies not just between the different species with the bare olivine 
substrate but also with ice and with the other adsorbed species. 
In fact, we considered all possible binding energies to describe the system. 
In addition to this, we also considered the freeze out of O and CO from the gas phase. 
These are the major improvement with respect to our previous model. \citet{Cup07} and \citet{Cup09}, also 
considered the formation of grain mantles. However, \citet{Cup07}, restricted themselves 
only to the formation of water and 
\citet{Cup09} restricted themselves only to the formation of methanol. They did not study the 
variation as function of initial abundances. We have extended the study of grain mantle  
by incorporating all the major grain surface species namely, H$_2$O, CH$_3$OH and 
CO$_2$ and CO which contains nearly 90 $\%$ of grain mantle composition and we have studied the variation 
of the abundances with respect to the initial conditions and cloud densities.

\section{Model}

\noindent In our calculation, only three gas phase species, namely, H, O and CO are allowed to accrete on a silicate grain 
surface. In total, on a grain surface, $10$ chemical reactions are considered which can occur among these three constituents via thermal hopping. 
H$_2$, O$_2$, H$_2$O, H$_2$CO, CH$_3$OH, and CO$_2$, as well as reactive intermediate species are formed. No gas phase chemistry 
is considered and the initial gas-phase concentrations of only CO and O are varied from Model to Model while the hydrogen concentration
is assumed to be constant. This will decrease the accretion rate of O and CO species with time. In other words, freeze out effect 
of these two molecules are taken into account. Because of this effect, the accretion rates of these molecules decrease and
the number of these species on the grain surface will also decrease. This will, in turn, 
result in a decrease in the formation rate of molecules which uses them.  

We consider conditions which are representatives of a dense cloud. In these clouds, most of the atomic hydrogen is converted
into molecular hydrogen and atomic carbon into CO via gas phase chemistry. Therefore, we have considered accretion of CO instead 
of atomic carbon. The abundances (cm$^{-3}$) of H, O, and CO are shown in Table 1. These values were obtained from steady-state gas-phase 
model runs at total hydrogen number densities (n$_h$) of $10^3$, $10^4$, and $10^5$ cm$^{-3}$ taken from \citet{Stan}. We designate 
them as `low', `intermediate' and `high' density cases. Furthermore, we have varied the initial abundances of O ($n_O$) and CO ($n_{CO}$) 
for intermediate density case to see the effects. We have shown earlier that the methanol and water can be efficiently formed in the clouds of intermediate 
density (\citet{Dasb}). We scan the entire parameter space to find favourable conditions for the production of water, methanol and carbon di-oxide.
We varied the O abundance between $7 \times 10^{-5}$ and $7 \times 10^{-4}$ and CO between $7.5 \times 10^{-6}$ and  $1.5 \times 10^{-4}$ 
with respect to the total hydrogen. 
SWAS findings of very low H$_2$O and O$_2$ put a serious concern over the gas phase O abundance. Gas and solid state molecules can account about 
55 $\%$ of the oxygen (\citet{Bergin00}). Therefore, remaining atomic oxygen is in the gas phase as was found toward a few sources
(\citep{Caux99, Lis01} and references therein). Another possibility is that there is a higher depletion of atomic oxygen
onto the dust grain. \citet{Bergin00}, found that using a C/O $>$ 0.9, they can re-produce the SWAS observations of H$_2$O and O$_2$ abundance
in star forming cores. We have chosen lower limit of O abundance such that we have C/O ratio close to 1. 
Atomic oxygen abundance as high 
as $7 \times 10^{-4}$ is used to mimic the conditions around solar neighborhood. We have chosen the CO initial abundance on the basis of 
various available studies \citep{Jogen, Wilson}. We have considered the grain temperature to be at $10$ K. In this temperature, only H, O and H$_2$ have 
mobility and CO is almost immobile, therefore we have neglected the hopping of CO thereby saving the computational time. We presented in this paper a 
few selected results out of our many model runs.

\section{Results}
We found that the formation of water, methanol and carbon dioxide on a grain surface strongly depends
on initial gas phase abundances of O and CO and on the cloud density. Initial abundance and cloud density dictate the 
relative accretion rates of O and CO thereby controlling the formation of these molecules.

\subsection{Variation of abundance due to initial abundance}

In Fig. 1(a-l), we show the variation of abundances of selected species as a function of time for different initial 
conditions and for intermediate cloud density, i.e., the total hydrogen number density is $10^4$ cm$^{-3}$. We vary the gas 
phase atomic abundance between $7 \times 10^{-5}$ and  $7 \times 10^{-4}$ and CO abundance between $7 \times 10^{-5}$ and 
$1.5 \times 10^{-5}$ with respect to the total hydrogen number density. Table 4, summarizes abundances of these molecules 
after two million years. In Fig. 1a, the initial O abundance of $7 \times 10^{-5}$ and 
CO abundance of $7.5 \times 10^{-5}$ were used. 
We find that under this condition there is more methanol than water. The reason is that the accretion rate of 
CO is little higher ($~ 7\%$) than O and it is mainly utilized to form methanol and CO$_2$ whereas, O
is utilized to form water, O$_2$ and CO$_2$. The next most abundant molecule is CO$_2$. It is mainly formed through 
reaction number $10$ in Table 3. We find that at around $2 \times 10^6$ years, the 
rate of formation of these molecules slowed down significantly, this 
is because we have considered the freeze out effects of CO and O. As the number densities of 
these two species go down in gas phase, their accretion rates also go down which results  in
a decrease in the rate of formation of other molecules. When O and CO 
are heavily depleted from the gas phase, we obtain a near steady state on the grain surface. 
In Figs. 1b and 1c,  we have decreased 
the initial CO abundance by a factor of half and one-fifth respectively 
for a fixed O abundance, which results in a decrease in methanol and an increase 
in water abundance. Water is accounted for nearly $60 \%$ and $80 \%$ of the grain 
mantle respectively. The methanol abundance is $52 \%$ and $20 \%$ with respect to 
water and CO$_2$ is $9 \%$ and $3 \%$ respectively (Table 4). 

At this stage, it is pertinent to ask that how close these results are to the observed values. 
We find that in almost all the cases methanol is over estimated and CO$_2$ is under estimated \citep{Kean}. 
In Fig. 1d, we increase the abundance of O.  
The net affect is a decrease of methanol and a small increase in the abundance of molecular oxygen. 
Once again, if we compare this set with the observed value \citep{Kean} we find that the abundance of water and CO$_2$ is close to observed values. However, 
the methanol is over-produced. In Fig. 1e, we decreased the abundance CO by half, which results in  a farther reduction of methanol. 
For this case, the abundance of methanol and water is in close agreement with the observed value in GL 7009S. However, abundance of 
CO$_2$ is nearly half of the observed abundance. In Fig. 1f, CO abundance is one-fifth and  $\sim 90 \%$ of the
 grain mantle is made up of 
water. It is understandable because with this initial condition, the accretion rate of O is nearly one order of magnitude greater 
than CO. Next we increase the initial O abundance by five times (Fig. 1g). Once again, the water is the most 
abundant species, methanol and CO$_2$ 
is within the observed limit but a substantial increase in the abundance of molecular oxygen is seen. 
Thus a high abundance of molecular oxygen could be found in a oxygen rich environment. In Figs. 1h and 1f, 
results are shown for half and one-fifth CO abundance. 
The abundance of CO in most of the cases is very low. This is due to the fact that 
CO in a grain is rapidly converted into methanol and CO$_2$. Finally, we have increased the 
O abundance $10$ times and found a substantial increases in molecular oxygen abundance (Fig. 1j, 1k and 1l).

In Fig. 2, a cross-sectional view of the mantle structure after one million year is shown.
The O abundance of $1.05 \times 10^{-4}$ and CO 
abundance of $1.5 \times 10^{-5}$ were used in the simulation. In this case,
water forms about $85 \%$ of the grain mantle, methanol about $13 \%$ and 
CO$_2$ about $3 \%$ of water. Total number of mono-layer formed is $ 85$. The species having less 
than $1\%$ surface coverage are not shown in the Figure.

In Fig. 3 and Fig. 4, we show the variation of final mantle abundance of selected species as a function of initial abundance of gas 
phase O and CO. This Figure is analogous to Fig. 1. In Fig. 1 we have shown as a function of time and in these two figures we have 
plotted abundances of selected species as a function of accreted gas phase O 
abundance and CO. In Fig. 3, each column represents a fixed value of CO: $7.5 \times 10^{-5}$ (1st column), $3.75 \times 10^{-5}$ (2nd column) 
and $1.5 \times 10^{-5}$ (3rd column) with respect to total hydrogen number density and initial gas phase O abundances ranging from 
$7 \times 10^{-5}$ to $7 \times 10^{-4}$ with respect to total hydrogen number density. 
Purpose of this plot is to get an idea of how absolute abundance changes with initial condition. 
With increasing gas phase O, production of oxygenated species 
like H$_2$O, O$_2$ and CO$_2$ increases whereas the production of methanol decreases. 
Grain phase CO also less populated with the increase in initial gas phase O as CO 
are mostly converted to CO$_2$. In Fig. 4, we choose initial  gas phase O to be $7 \times 10^{-4} $(1st column), 
$1.05 \times 10^{-4} $(2nd column) and $7 \times 10^{-5} $ (3rd column) and show the final abundances of various species at the 
grain mantle for different initial gas phase CO abundances ranging from $1.5 \times 10^{-5}$ to $3.5 \times 10^{-4}$. It is clear 
from the Figure that as the initial gas phase CO abundance increases, methanol production increases. Due to the enhance hydrogenation of CO 
for the production of methanol, production of H$_2$O decreases which result some O free enough to react with other O or CO to produce O$_2$ or 
CO$_2$. So O$_2$ and CO$_2$ also increases with the increase in the gas phase CO abundances.

\subsubsection{Grain mantle thickness}
In Fig. 5, we show the time evolution of the number of mono-layers. Time evolution of the number of layer is shown for initial gas phase
CO=$7.5\times 10^{-5}$, $3.75\times 10^{-5}$ and $1.5\times 10^{-5} $ and for different gas phase O noted on the right side of the box.
It is found that the mantle thickness is a function of initial gas phase abundance and thickness varies between $60$ to $500$
mono-layers. For higher initial abundance, accretion rate is high, which results higher mantle thickness on the grain because now grain
has more matter to process.

\subsubsection{Parameter space and zone of interest}

By varying the initial abundances of O and CO we find that in almost all the cases, water is the most abundant species. 
The next most abundant 
species is the methanol which is generally contrary to the various observational results. Next most abundant species 
is CO$_2$ which is generally 
close to the observed values. This prompted us to run our code for a wide set of parameters to find out parameter space in which these molecules 
are produced in the observed range. Results are presented in Figs. 6a, 6b and 6c. We plot water, methanol and CO$_2$ coverages on the grain mantle as a 
function of initial abundances of O and CO and marked the zone in which these molecules are within the observed limit. We call this as a 
favourable zone. Thus the favourable zone excludes abundances which are over produced or less produced with respect to the observed limit. Observed 
limits are selected using \cite{Kean}. To find out the favourable zone we considered those cases for which at least $80$ mono-layers are formed, 
water abundance is around $10^{-4}$, methanol 
abundance is between  $2\%-30\%$ of the solid state water and CO$_2$ abundance between 2\% and 20\% of the solid state water. We find that the water 
abundance is close to observed value when CO abundance is less than $5 \times 10^{-5}$ and O is between $3.5 \times 10^{-5}$ and $7 \times 10^{-4}$. 
Methanol is close to observed value when CO abundance is between $7.5 \times 10^{-6}$ and $1.5 \times 10^{-4}$ and O is between $3.5 \times 10^{-5}$ and 
$7 \times 10^{-4}$. Similarly, with CO$_2$, this zone for CO is between $1.5 \times 10^{-5}$ and $7.5 \times 10^{-5}$ and O between $3.5 \times 10^{-5}$ 
and $7 \times 10^{-4}$. In Fig. 6d, we have superimposed favourable zones from all the three species to see whether there is any region  which is common  
to all the three favourable zones. It is clearly seen that only in a very narrow region of the parameter space, 
the observed abundance could be roughly reproduced.

\subsubsection{Effect of the activation energy}

We have performed another exercise by increasing the activation energy of methanol formation to check if this could reduce methanol formation. 
We increase the activation barrier energy by a factor of $1.5$, $2$ and $2.5$ times with 
respect to the activation energy barrier used by several authors. An increase in activation 
energy up to two times has almost no effect on methanol production. However, when the 
activation energy is $5000$ K ($2.5$ times higher) we find a dramatic reduction 
of methanol formation (Table 5). This also reduces CO$_2$ production 
because the efficient formation route for CO$_2$ in our method is found to be the reaction no. $10$ of Table 3. 
Under this condition, a lot of unused CO remains on the grain and a enhanced production of CO is observed.

\subsection{Variation of abundance with cloud density}
In Fig. 7, we plot the abundance of selected species as functions
of time for various cloud density for initial abundance shown in Table 1 and
result is tabulated in Table 6. In the low density case only water and methanol is produced efficiently and 
the abundance is also not close to the observed value.
In the intermediate case, the most abundant species is methanol rather than water. 
A reasonably good amount of CO$_2$ is also produced. Finally, in the high case, everything
is heavily produced, especially those requiring CO in the reaction scheme. 
This is because under this condition hydrogen supply to saturate CO is
insufficient. Our general conclusion is that by varying cloud density and using standard formation 
route of water and methanol it is difficult to reproduce the observed abundances for these molecules.

\subsection{Collapsing cloud}

In a realistic molecular cloud, the density will change with time due to a gradual collapse of the cloud. 
To understand how the abundance of various species changes with density we now consider a 
cloud which collapses with time. However, since coupling of the grain chemistry and a gas chemistry in presence of 
a collapsing cloud is a very difficult task 
we made some simplification. For simplicity, we use the radial density distribution from \cite{Dasa}. 
where a spherically symmetric, self gravitating,  isothermal, collapsing cloud was considered. Initially this cloud is assumed to have a
negligible amount of mass. The outer boundary is at 1 parsec and the inner boundary is at 10$^{-4}$ parsec. The matter which 
crosses the inner boundary is assumed to form the core. Matter is injected at a constant rate at the outer boundary. In \cite{Dasa}, the  cloud was 
divided into $100$ logarithmically equal spaced grids.  Here, we consider only 5 shells, each containing 20 grids. We take volume 
average of number densities in each shell to have an average density. 
We have taken the initial number density of various species to be same as it was for $10^3$cm$^{-3}$ 
(low density, Table 1). Time evolution of the number density for a collapsing cloud is shown in Fig. 8a. 
The system is evolved with time due to collapse and incorporating the 
freeze-out and desorption effects. Abundance of a few selected species are shown in Fig. 8b where the 
five data points from the five shells could be seen.
Methanol is varying between 60 \% and 100 \% of water, a signature we observed when cloud density was varied. Thus methanol 
is always over-produced. Deep inside the cloud, the methanol abundance begins to drop due to the 
unavailability of H atoms. CO$_2$ is produced within the observed limit 
between outer shell to middle shells. However, as we go deep inside the cloud CO$_2$ is heavily produced. 
Very little O$_2$ is produced up to middle shells but a very good amount of O$_2$ is  
produced deep inside the cloud. We found that as the density increases, H atom is rapidly used up due to hydrogenation
reaction. As a result, we found H$_2$O and CH$_3$OH is produced very efficiently in the outer shells.
In the process, as the cloud collapses, the relative numbers of CO and O go up with respect to H atom.
This then favours the formation of CO$_2$ and O$_2$ in the inner shells. This resembles the abundances 
we obtained by varying 
cloud density. Thus once again we can conclude that 
the final abundances of the interstellar species are strongly dependent on the relative initial abundances of 
its constituents.

\subsection{Comparison with previous works}

In \citet{Dasb}, water and methanol formations in a mono-layer on grain surfaces were studied, where freeze-out effects of 
gas phase species were not considered. In the present paper, we study the evolution of the mantle up
to several tens of layers in presence of freeze-out effects. We also
identify the parameters space in which the produced water, methanol and carbon dioxide
are comparable to what are observed.

We have compared our results with that obtained from the rate equation method of \citet{Hase}. This we present in Fig. 9.
We put `(m)' for Monte-carlo method and `(r)' for rate equation method.
We found that the difference between the Monte-Carlo method and the rate equation method is
negligible for H$_2$O and CH$_3$OH. However, for the species with smaller abundances, the differences are significant.
For instance, our production of CO$_2$ on grains via Eq. 10 of Table 3 is more efficient (than that
via Eq. 9) where HCO is a reactant. In the rate equation method, it  is likely that CO$_2$ would be estimated
incorrectly due to very low abundance of HCO on the grain.
This is reflected in Fig. 9 also. Discrepancies in other species can be similarly explained.
The species with higher abundances are not affected because the rate equation method is valid in this regime.
In this regard we can compare our results with some other previous works. 
\citet{Stan04} considered a gas-grain coupled
network to study the chemical evolution of a static molecular cloud. Similar to our approach, their model
also seems to be efficient for the production of methanol. Major differences between our results are that
at the end we are obtaining significant amount of solid CO and CO$_2$, where 
their result shows virtually no solid CO and little solid CO$_2$. In \citet{Ruff00}, 
a complete network of
surface reactions were considered to study the quiescent source in front of the field star 
Elias 16, they were able to reproduce the 
CO abundances and upper limit to the Methanol abundances by considering much lower 
diffusion rates than the diffusion rates used here. However, all of their model were 
unable to produce significant about of CO$_2$.

While comparing our results with other multilayer methods, we note that,
\citet{Cup07}, studied the formation of water on grain surface for diffuse, translucent and dense clouds. 
The most striking difference of \citet{Cup07} with the present work is that they did not include methanol formation at all.
In our case, the H atom is partially used up to form methanol. Furthermore, in the absence of a few important 
species on the grain, such as methanol, carbon monoxide etc.
they always found that water dominates the grain mantle. On the contrary, we found that for 
C/O $\ge$ 1, methanol is produced more than water. Recently, \citet{Cup09}, studied the surface formation 
of CH$_3$OH and H$_2$CO from precursor CO using the continuous-time, random walk Monte-Carlo method. They found the
formation of both the species to be very efficient and the efficiency 
depends mainly on the grain temperature and the abundance ratio of H and CO in the gas phase. They also found
that the freeze-out of CO favours more complete hydrogenation of CO to form CH$_3$OH.
We found, indeed the methanol is formed very efficiently. However, in most
of the cases formaldehyde in our case is less than $1 \%$ while it is somewhat higher 
in \citet{Cup09}. Another difference with \citet{Cup09} is that we have considered the
accretion of O also on the grain surface and thus we find that some fraction of the mantle is
covered with O2 and CO2. Varying the initial cloud composition(by changing the initial gas phase O and CO), we
have noticed that in a narrow region of parameter space(Fig. 6d) our model is producing H$_2$O, CO$_2$, and 
CH$_3$OH within the observed limit. In the other regions of the parameter space we are getting
an anti-correlation between the production of CH3OH, CO$_2$ with H$_2$O.

\section{Conclusions}
In this paper, we have carried out numerical simulations using continuous-time random walk Monte-Carlo method to investigate 
formation of water, methanol and CO$_2$ as a function of the cloud density and the initial abundances of CO and O. 
We found that when the accretion rate of CO is more than O, methanol 
is the most abundant species on the grain surface. An increase in the accretion rate of O over CO, methanol abundance gradually goes down. When the accretion 
rate of O is $1.4$ times that of the CO accretion rate, the water abundance is more than that of methanol, 
i.e, a cross over is seen. One has to increase O/CO by a factor of $3$ to get a methanol 
abundance which is $ \sim 30 \%$, the maximum abundance of methanol seen in the molecular clouds \citep{Kean}. When we increase 
the accretion rate of O further, methanol starts to get reduced and this trend continues.
When we increase the ration O/CO by a factor of more than $10$, 
although water is the most abundant molecule on the grain surface, we observed a serge in O$_2$ abundance. 
We scanned the entire parameter space and found that in the clouds having 
$O/CO < 3$,  methanol is overproduced. This criteria is true only  
when there is no depletion, because we have more O than CO to start with. 
Once the accretion process sets in, the gas phase species will get depleted on to the 
grain and we will have clouds with different C/O ratio. The cloud conditions in which 
the C/O ratio is less than $0.33$, water is found to be produced very efficiently on the grain surface.
We concede that matching CO$_2$ abundance with the observational results still remains a problem.  We found 
that CO$_2$ is mainly formed through reaction no. $10$ in Table 3, i.e., it requires HCO to form. 
Therefore, conditions which lead to decrease in methanol formation 
also decrease  the formation of CO$_2$. We have also checked the effects 
of increasing activation barrier energy for methanol formation, but the result is not improved as far as 
matching with observed abundances is concerned. We also failed to match with observed abundances by changing
the cloud density only. However, when we super-impose the
favourable formation zones of H$_2$O, CH$_3$OH and CO$_2$ we do obtain a narrow region in which all these molecules are 
produced within the observed limit. {\it To have more realistic estimation about the abundances of 
various interstellar species 
in the different regions of the molecular cloud, we have considered a collapsing cloud and have coupled 
the time evolution of density distribution with our Monte Carlo approach of the grain chemistry. We have noticed that
throughout the cloud CH$_3$OH is always overproduced. Deep inside the 
cloud CO$_2$ and O$_2$ are heavily produced whereas at the outer edge their production 
gradually decreases.
}

\section{Acknowledgments}
A. Das wishes to acknowledge the hospitality of S. N. Bose National Centre for Basics Sciences when the paper was written.

\clearpage


\begin{table}
\centering
\caption{Initial Gas-phase abundances used}
\begin{tabular}{|c|c|c|c|}
\hline
Species & high ($cm^{-3}$) & intermediate($cm^{-3}$) & low ($cm^{-3}$) \\
\hline
H&1.10 &1.15 &1.15\\
\hline
O& 7.0 & 0.75 & 0.09\\
\hline
CO&7.5 &0.75 &0.075\\
\hline
\end{tabular}
\label{table-1}
\end{table}

\begin{table}
\centering
\caption{Energy barriers in degree Kelvin}
\begin{tabular}{|c|c|c|c|c|c|c|c|c|c|c|c|c|c|c|}
\hline
Species & \multicolumn{13}{|c|}{Substrate}\\
\hline

& Silicate &H&$\mathrm{H_2}$&O&$\mathrm{O_2}$&OH&$\mathrm{H_2O}$&CO&HCO&$\mathrm{H_2CO}$&$\mathrm{H_3CO}$&$\mathrm{CH_3OH}$&$\mathrm{CO_2}$\\
\hline
&$E_D$&$E_D$&$E_D$&$E_D$&$E_D$&$E_D$&$E_D$&$E_D$&$E_D$&$E_D$&$E_D$&$E_D$&$E_D$\\
\hline
H&350&350&45&350&45&350&650&350&350&350&350&350&350\\
\hline
$\mathrm{H_2}$&450&30&23&30&30&30&440&450&450&450&450&450&450\\
\hline
O&800&480&55&480&55&55&800&480&480&480&480&480&480\\
\hline
$\mathrm{O_2}$&1210&69&69&69&69&1000&1210&1210&1210&1210&1210&1210&1210\\
\hline
OH&1260&1260&240&240&240&240&3500&1260&1260&1260&1260&1260&1260\\
\hline
$\mathrm{H_2O}$&1860&390&390&390&390&390&5640&1860&1860&1860&1860&1860&1860\\
\hline
CO&1210&1210&1210&1210&1210&1210&1210&1210&1210&1210&1210&1210&1210\\
\hline
HCO&1510&1510&1510&1510&1510&1510&1510&1510&1510&1510&1510&1510&1510\\
\hline
$\mathrm{H_2CO}$&1760&1760&1760&1760&1760&1760&1760&1760&1760&1760&1760&1760&1760\\
\hline
$\mathrm{H_3CO}$&2170&2170&2170&2170&2170&2170&2170&2170&2170&2170&2170&2170&2170\\
\hline
$\mathrm{CH_3OH}$&2060&2060&2060&2060&2060&2060&2060&2060&2060&2060&2060&2060&2060\\
\hline
$\mathrm{CO_2}$&2500&2500&2500&2500&2500&2500&2500&2500&2500&2500&2500&2500&2500\\
\hline
\end{tabular}
\vskip 0.2cm
\scriptsize{For the all species $E_b=0.3\ E_D$ is used except the atomic hydrogen where
$E_b=0.2857 \ E_D$ is used. For the reference please see the text.}
\label{table-2}
\end{table}

\begin{table}
\centering
\caption{Surface Reactions in the H, O, and CO model}
\begin{tabular}{|l|l|l|}
\hline
Number & Reactions & E$_a$(K)              \\
\hline
1  & H+H $\rightarrow$ H$_2$          &      \\
2  & H+O $\rightarrow$ OH             &      \\
3  & H+OH $\rightarrow$ H$_2$O        &      \\
4  & H+CO $\rightarrow$ HCO           & 2000 \\
5  & H+HCO $\rightarrow$ H$_2$CO      &      \\
6  & H+H$_2$CO $\rightarrow$ H$_3$CO  & 2000 \\
7  & H+H$_3$CO $\rightarrow$ CH$_3$OH &      \\
8  & O+O $\rightarrow$ O$_2$          &      \\
9 & O+CO $\rightarrow$ CO$_2$        & 1000 \\
10 & O+HCO $\rightarrow$ CO$_2$+H     &      \\
\hline
\end{tabular}
\label{table-3}
\end{table}

\begin{table}
\footnotesize
\caption{Relative abundances (absolute abundance) of the ice species after two million years}
\begin{tabular}{|c||c|l|l|l|}
\hline
Initial gas phase &Species&Initial gas phase&Initial gas phase&Initial gas phase\\
 Oxygen&&CO=7.5(-5)&CO=3.75(-5)
&CO=1.5(-5)\\
\hline
\hline
&$\mathrm{H_2O}$&100 [5.3(-5)]&100 [6(-5)]&100 [6.4(-5)]\\
\cline{2-5}
&CO&3\hspace{2 sp}    [1.6(-6)]&1 [6.1(-7)]&0.5 [2.9(-7)]\\
\cline{2-5}
7.0(-5)&$\mathrm{H_2CO}$&0.2\hspace{1 mm} [8.8(-8)]&0.05 [3(-8)]&0.02 [1.1(-8)]\\
\cline{2-5}
&$\mathrm{CH_3OH}$&116\hspace{0.5 mm} [6.1(-5)]&52.3 [3.2(-5)]&20 [1.3(-5)]\\
\cline{2-5}
&$\mathrm{CO_2}$&22.7 [1.2(-5)]&8.6 [5.2(-6)]&2.8 [1.8(-6)]\\
\cline{2-5}
&$\mathrm{O_2}$&5 \hspace{2 mm} [2.6(-6)]&3.6 [2.2(-6)]&3 [2(-6)]\\
\hline
\hline

&$\mathrm{H_2O}$&100 [7.5(-5)]&100 [8.8(-5)]&100 [9.4(-5)]\\
\cline{2-5}
&CO&0.8 [5.8(-7)]&0.4 [3.7(-7)]&0.3 [2.4(-7)]\\
\cline{2-5}
1.05(-4)&$\mathrm{H_2CO}$&0.05 [4(-8)]&0.01 [1.2(-8)]&0.01 [7(-9)]\\
\cline{2-5}
&$\mathrm{CH_3OH}$&70.3 [5.3(-5)]&33.4 [2.9(-5)]&13 [1.2(-5)]\\
\cline{2-5}
&$\mathrm{CO_2}$&20.7 [1.5(-5)]&7.8 [6.8(-6)]&2.5 [2.4(-6)]\\
\cline{2-5}
&$\mathrm{O_2}$&7 [5.3(-6)]&5.4 [4.7(-6)]&4.6 [4.4(-6)]\\
\hline
\hline

&$\mathrm{H_2O}$&100 [2.1(-4)]&100 [2.3(-4)]&100 [2.1(-4)]\\
\cline{2-5}
&CO&0.1 [2.1(-7)]&0.02 [4.5(-8)]&0.02 [3.5(-8)]\\
\cline{2-5}
3.5(-4)&$\mathrm{H_2CO}$&0.01 [1.9(-8)]&0.0 [4.8(-9)]&0.0 [1.6(-9)]\\
\cline{2-5}
&$\mathrm{CH_3OH}$&19.4 [4.1(-5)]&9.5 [2.2(-5)]&3.3 [6.8(-6)]\\
\cline{2-5}
&$\mathrm{CO_2}$&14.3 [3(-5)]&6.2 [1.4(-5)]&2.5 [5.2(-6)]\\
\cline{2-5}
&$\mathrm{O_2}$&24.2 [5.1(-5)]&21.7 [5(-5)]&23.7 [4.9(-5)]\\
\hline
\hline

&$\mathrm{H_2O}$&100 [2.4(-4)]&100 [3.4(-4)]&100 [3.6(-4)]\\
\cline{2-5}
&CO&0.03 [8.4(-8)]&0.02 [6.5(-8)]&0.00 [3.7(-8)]\\
\cline{2-5}
7.0(-4)&$\mathrm{H_2CO}$&0.00 [1.2(-8)]&0.00 [4.8(-9)]&0.00 [1.1(-9)]\\
\cline{2-5}
&$\mathrm{CH_3OH}$&8.5 [2.2(-5)]&4.9 [1.7(-5)]&2 [7.1(-6)]\\
\cline{2-5}
&$\mathrm{CO_2}$&15 [3.8(-5)]&5.7 [2(-5)]&2.1 [7.6(-6)]\\
\cline{2-5}
&$\mathrm{O_2}$&63 [1.6(-4)]&48.6 [1.7(-4)]&46.8 [1.7(-4)]\\
\hline
\hline

\end{tabular}
\label{table-4}
\end{table}



\begin{table}
\footnotesize
\caption{Relative abundances of the ice species by assuming different activation 
barrier energy for  H+CO and H+H$_2$CO reactions in the intermediate initial 
abundance case}
\begin{tabular}{|c||c|c|c|c|}
\hline
Species&2000&3000&4000&5000\\
\hline
\hline
$\mathrm{H_2O}$&100&100&100&100\\
\hline
CO&3&2.2&15.6&85.3\\
\hline
$\mathrm{H_2CO}$&0.2&0.06&0.5&0.2\\
\hline
$\mathrm{CH_3OH}$&116&111&95&19.4\\
\hline
$\mathrm{CO_2}$&22.7&19&7&1.5\\
\hline
$\mathrm{O_2}$&5&5.4&5&5.3\\
\hline
\hline
\end{tabular}
\label{table-5}
\end{table}


\begin{table}
\footnotesize
\caption{Relative abundances of the ice species in different regions of the interstellar cloud} 
\begin{tabular}{|c|c|c|c|}
\hline
Species&Low&Intermediate&High\\
\hline
\hline
$\mathrm{H_2O}$&100&100&100\\
CO&0.7&3&160\\
$\mathrm{H_2CO}$&0.02&0.2&13\\
$\mathrm{CH_3OH}$&68&116&146\\
$\mathrm{CO_2}$&2.5&23&143\\
$\mathrm{O_2}$&0.5&5&96\\
\hline
\hline
\end{tabular}
\label{table-6}
\end{table}

\begin {figure}
\vskip 2.0cm
\centering{
\includegraphics[width=18cm]{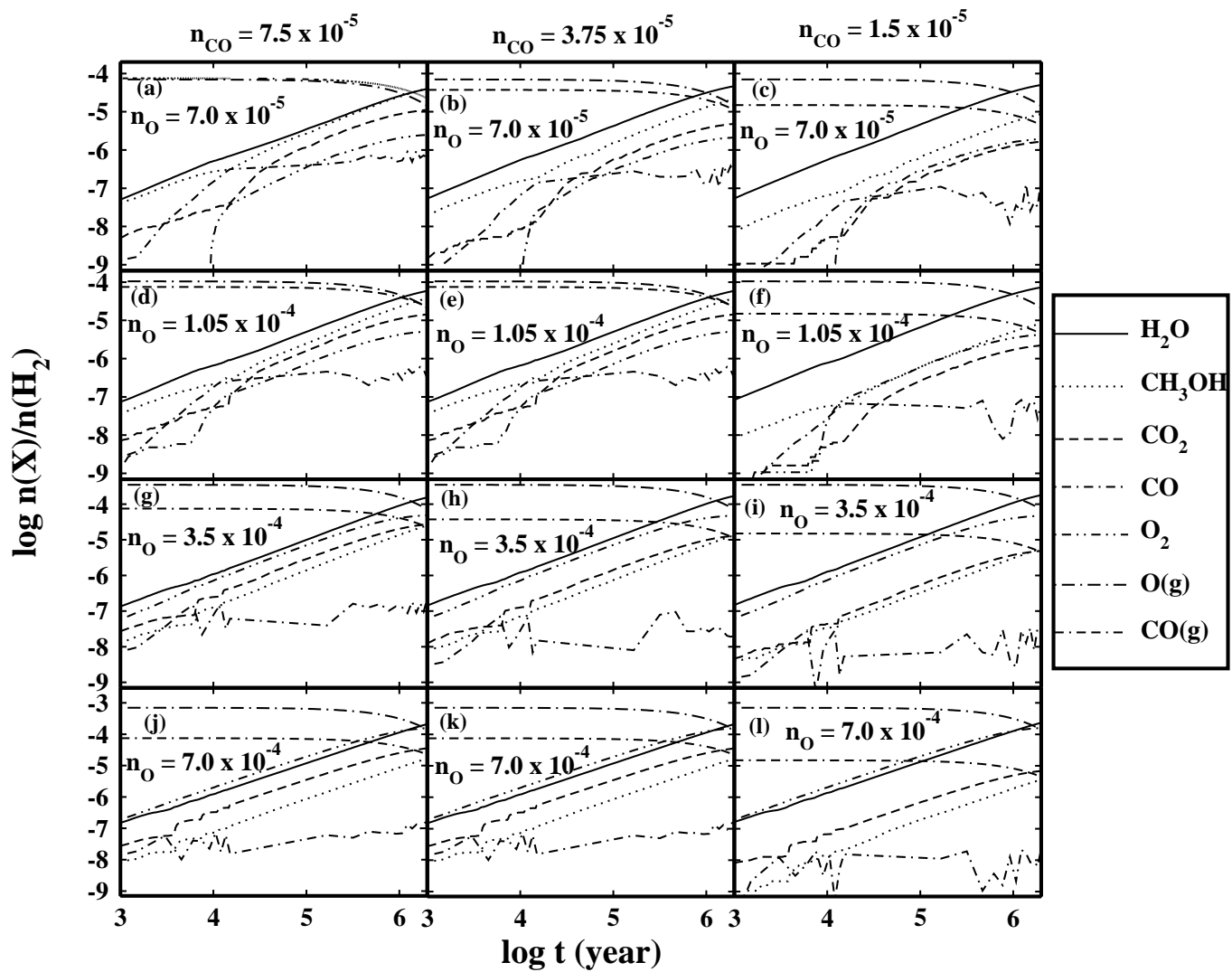}}
\caption{
Time variation of major grain surface species for different initial abundance
of carbon monoxide and oxygen. Chosen CO abundances are $7.5 \times 10^{-5}$ (1$^{st}$ column), $3.75 \times 10^{-5}$ (2$^{nd}$ column) 
and $1.5 \times 10^{-5}$ (3$^{rd}$ column).
}
\label{fig-1}
\end {figure}

\begin{figure}
\vskip 1cm
\centering{
\includegraphics[width=10cm]{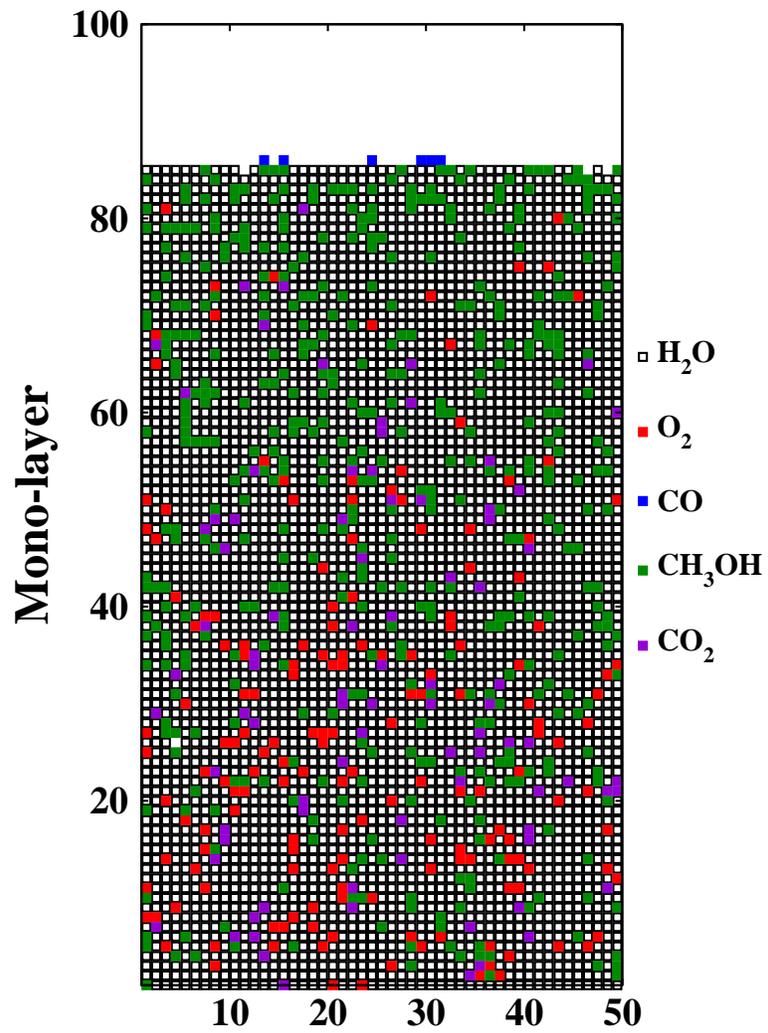}}
\caption{A cross-sectional view of the grain mantle after one million years for initial O and CO abundance of $1.05 \times 10^{-4}$ and $1.5 \times 10^{-5}$
respectively. }
\label{fig-2}
\end{figure}

\begin{figure}
\vskip 1.5cm
\centering{
\includegraphics[width=16cm]{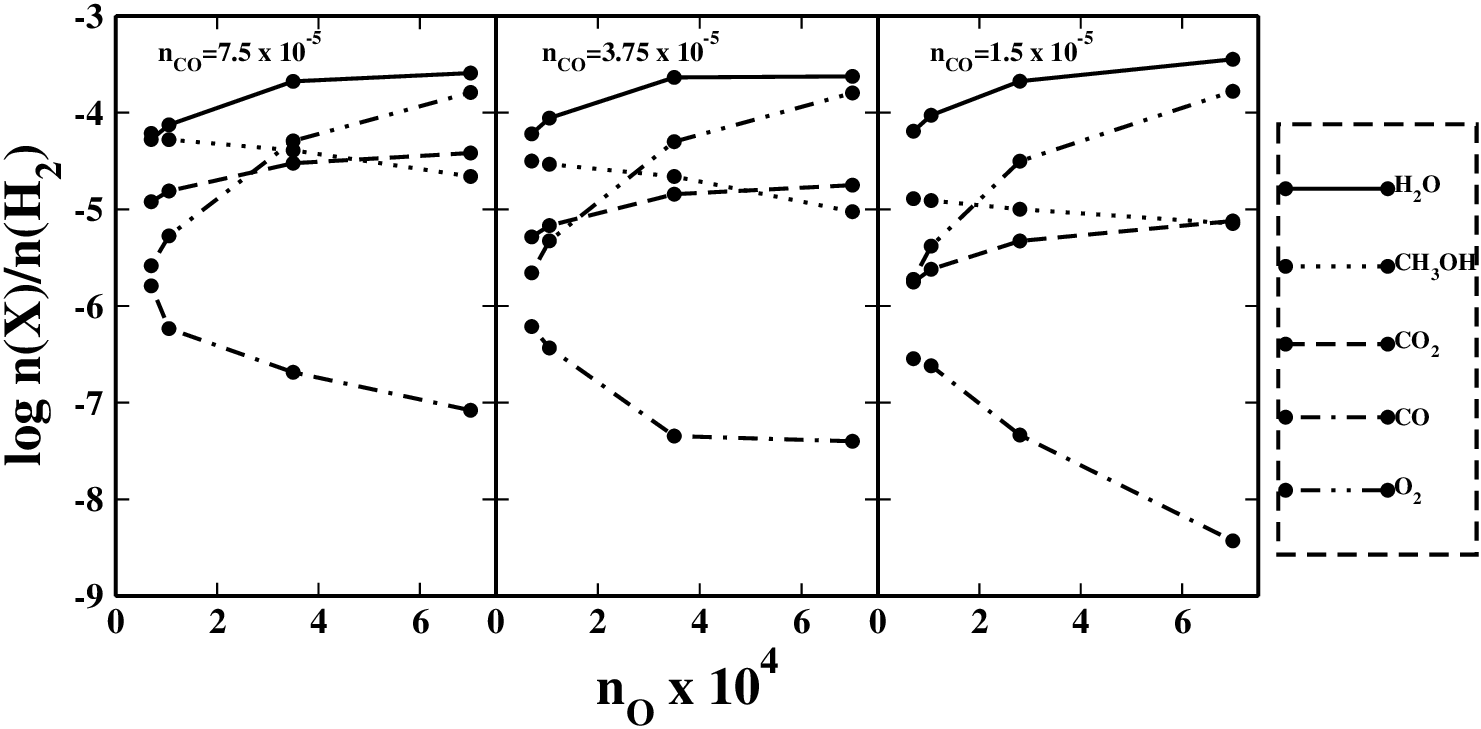}}
\caption{Variation of the final mantle abundance of the selected species as a function of initial abundance of gas
phase O.}
\label{fig-3}
\end {figure}

\begin{figure}
\vskip 2cm
\centering{
\includegraphics[width=16cm]{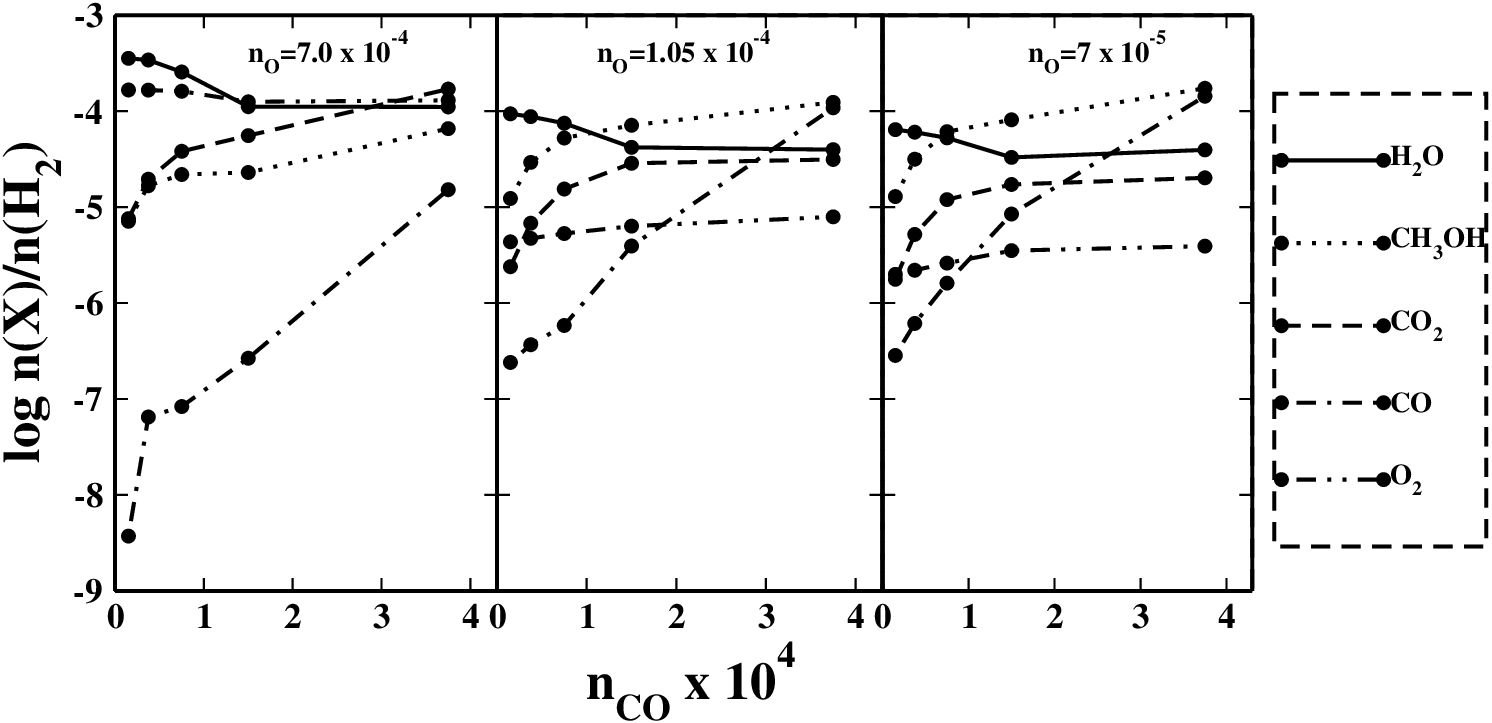}}
\caption{Variation of the final mantle abundance of selected species as a function of initial abundance of gas
phase CO.}
\label{fig-4}
\end {figure}

\begin{figure}
\vskip 2.5cm
\centering{
\includegraphics[width=12cm]{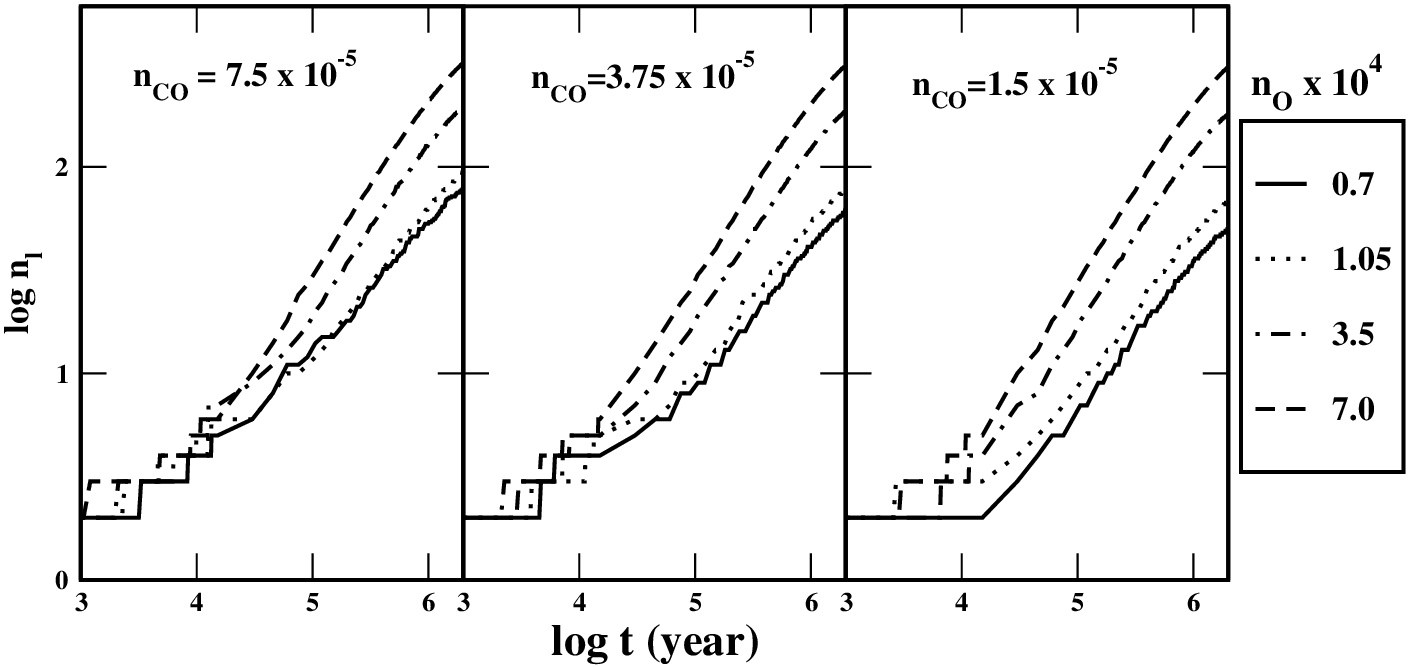}}
\caption{Time Variation of grain mantle thickness (in mono-layers) as function of initial O and CO abundance.}
\label{fig-7}
\end{figure}

\begin {figure}
\vskip 1.0cm
\centering{
\includegraphics[width=18cm]{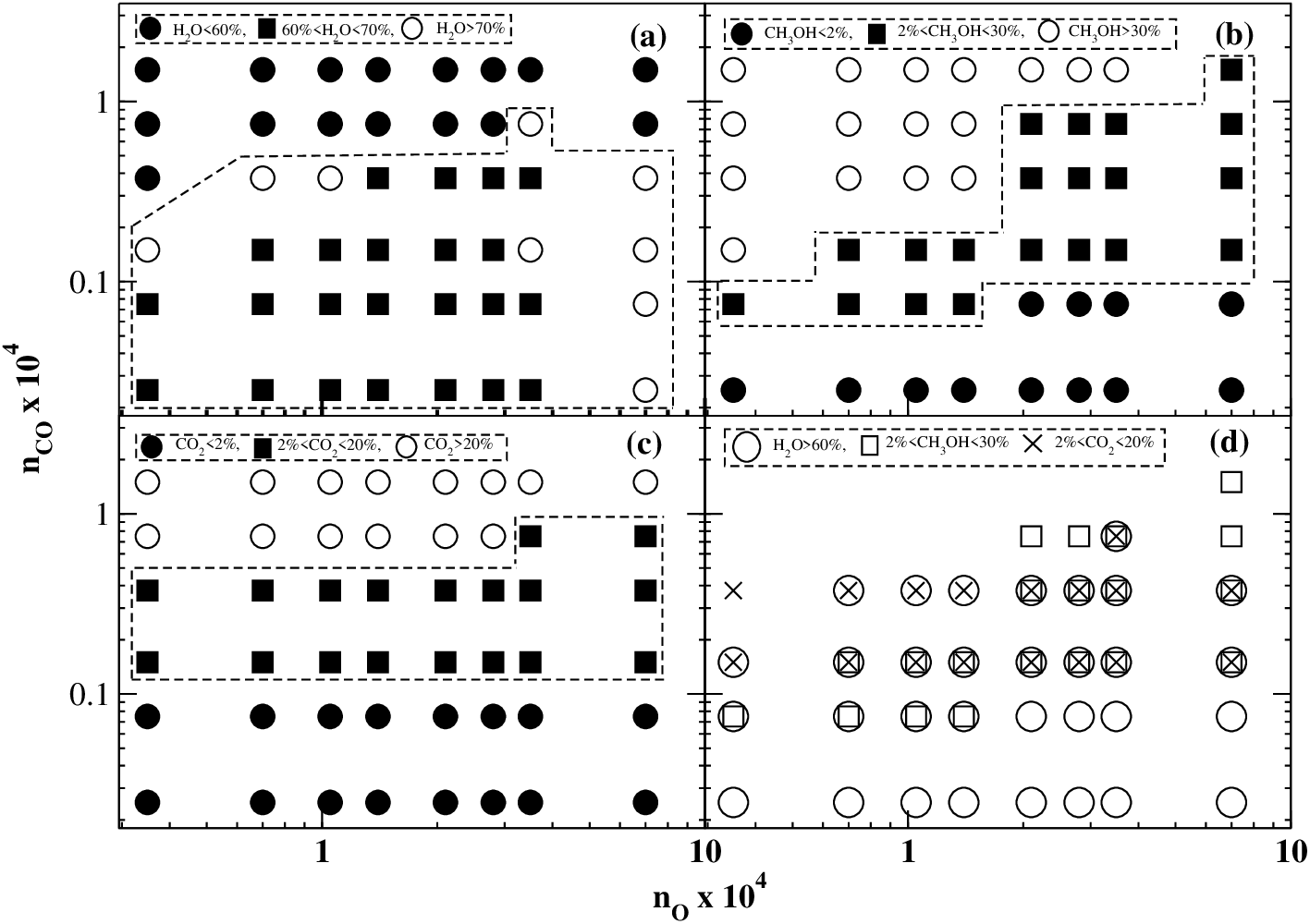}}
\vskip 1cm
\caption{
The parameter space in which formations of water, methanol and CO$_2$ are studied. The regions in which these molecules are 
produced {\bf within the observed limits} (favourable zones) is marked on the parameter space. 
(d) Favourable zone is the common zone when (a-c) are superimposed.
}
\label{fig-5}
\end {figure}

\begin {figure}
\vskip 0.5cm
\centering{
\includegraphics[width=12cm]{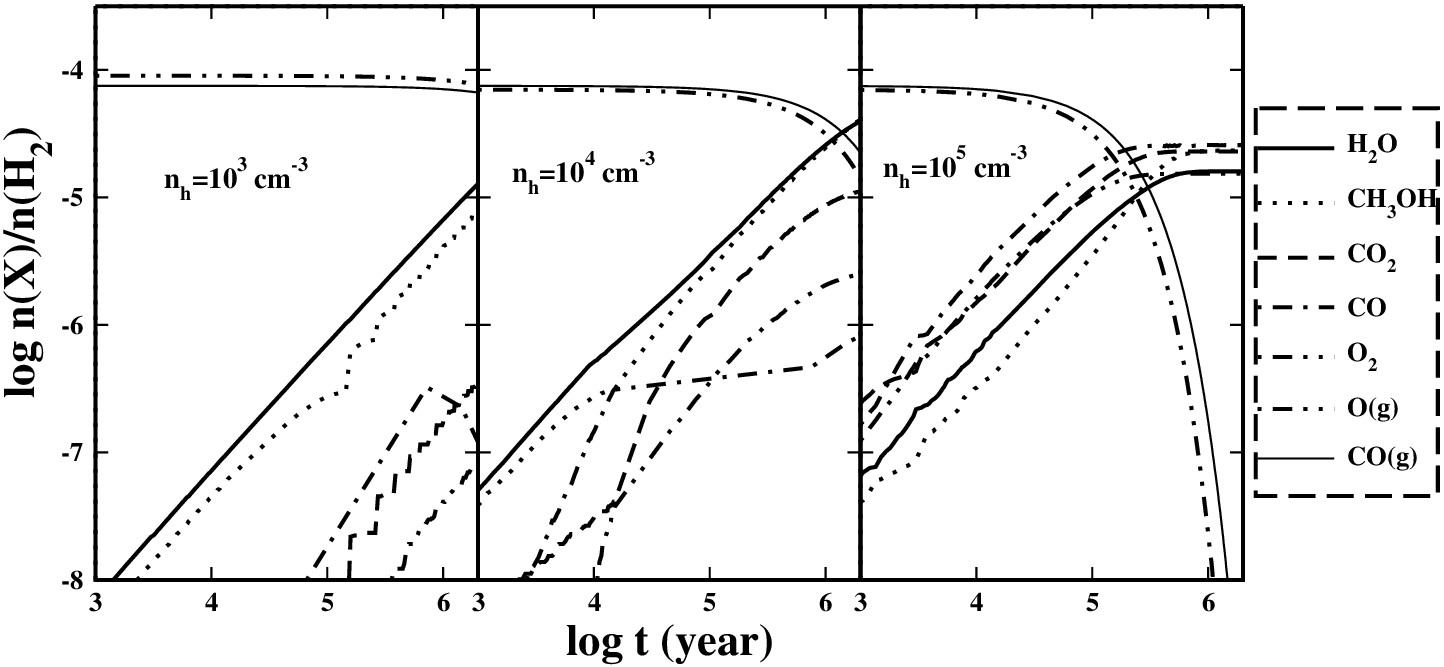}}
\vskip 1cm
\caption{Time evolution of selected species for various cloud densities.}
\label{fig-6}
\end {figure}

\clearpage
\begin {figure}
\vskip 0.5cm
\centering{
\includegraphics[width=12cm]{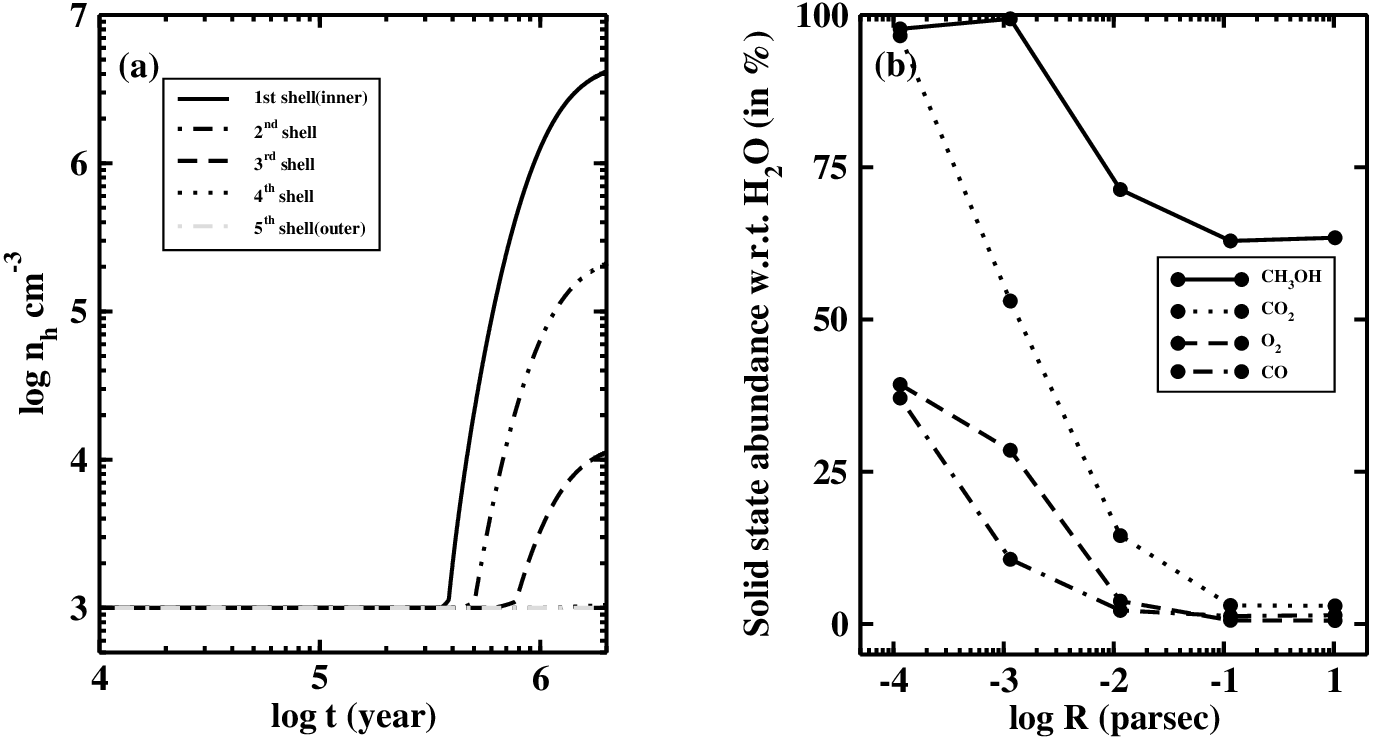}}
\vskip 1cm
\caption{Time evolution of (a)gas number density and (b)final abundances of some 
selected species in different region of a collapsing cloud.}
\label{fig-6B}
\end {figure}
\clearpage

\begin{figure}
\centering{
\includegraphics[width=12cm]{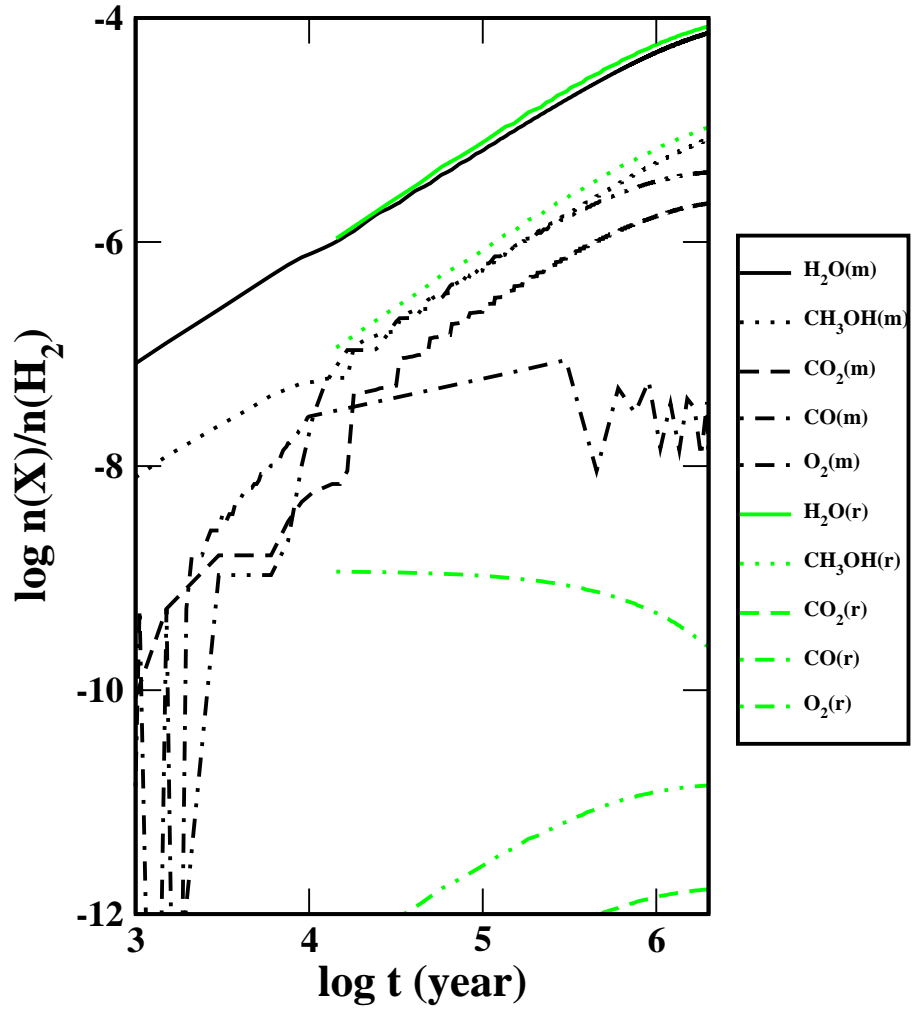}}
\caption{Comparison of results from Monte Carlo simulation (m) and rate equation (r) methods.}
\label{fig-8}
\end{figure}


\end{document}